\documentclass[tighten,iop]{emulateapj}

\usepackage[english]{babel}
\usepackage{amsmath}
\usepackage{amsfonts}
\usepackage{amssymb}
\usepackage{graphicx}
\usepackage[dvips]{color} 

\usepackage{natbib}

\shorttitle{3D Boltzmann-Hydrodynamical Code: II Moving-Mesh on PNS}
\shortauthors{Nagakura et al.}

\begin{document}


\title{Three-Dimensional Boltzmann-Hydro Code for core-collapse in massive stars

 II. The implementation of moving-mesh for neutron star kicks}

\author{Hiroki Nagakura$^{1,2}$, Wakana Iwakami$^{2,3}$, Shun Furusawa$^{4}$, Kohsuke Sumiyoshi$^{5}$, Shoichi Yamada$^{3,6}$, \\ Hideo Matsufuru$^{7}$ and Akira Imakura$^{8}$
}

\address{$^1$TAPIR, Walter Burke Institute for Theoretical Physics, Mailcode 350-17, California Institute of Technology, Pasadena, CA 91125, USA}

\address{$^2$Yukawa Institute for Theoretical Physics, Kyoto
  University, Oiwake-cho, Kitashirakawa, Sakyo-ku, Kyoto, 606-8502,
  Japan}

\address{$^3$Advanced Research Institute for Science \&
Engineering, Waseda University, 3-4-1 Okubo,
Shinjuku, Tokyo 169-8555, Japan}

\address{$^4$Center for Computational Astrophysics, 
National Astronimical Observatory of Japan, Mitaka, tokyo 181-8588, Japan}

\address{$^5$Numazu College of Technology, Ooka 3600, Numazu, Shizuoka 410-8501, Japan}

\address{$^6$Department of Science and Engineering, Waseda
  University, 3-4-1 Okubo, Shinjuku, Tokyo 169-8555, Japan}

\address{$^7$High Energy Accelerator Research Organization, 1-1 Oho, Tsukuba, Ibaraki 308-0801, Japan}

\address{$^8$University of Tsukuba, 1-1-1, Tennodai Tsukuba, Ibaraki 305-8577, Japan}

\begin{abstract}
We present a newly developed moving-mesh technique for the multi-dimensional Boltzmann-Hydro code for the simulation of core-collapse supernovae (CCSNe). What makes this technique different from others is the fact that it treats not only hydrodynamics but also neutrino transfer in the language of the 3+1 formalism of general relativity (GR), making use of the shift vector to specify the time evolution of the coordinate system. This means that the transport part of our code is essentially general relativistic although in this paper it is applied only to the moving curvilinear coordinates in the flat Minknowski spacetime, since the gravity part is still Newtonian. The numerical aspect of the implementation is also described in detail. Employing the axisymmetric two-dimensional version of the code, we conduct two test computations: oscillations and runaways of proto-neutron star (PNS). We show that our new method works fine, tracking the motions of PNS correctly. We believe that this is a major advancement toward the realistic simulation of CCSNe.
\end{abstract}



\section{Introduction}\label{sec:intro}

There are a number of observational and theoretical indications that the inner engine of core collapse supernovae (CCSNe) is highly non-spherical. Recent observational developments include three-dimensional direct mapping of $^{44}{\rm Ti}$ by NuStar, which reveals that the inner parts of the ejecta of SN~1987A and Cas~A have experienced large-scale mixing and convection \citep{2014Natur.506..339G,2015Sci...348..670B}. This is consistent with the earlier evidence from polarimetric observations that the explosion is not spherical in general and becomes more so as ones sees deeper inside (see e.g., \citet{2008ARA&A..46..433W} and references therein). High spatial velocities of pulsars are suggested to result from the recoil in asymmetric explosions \citep{2009A&A...497..423M}.

 On the theoretical side, there have been various mechanisms proposed as the cause for these multi-dimensional features, which may be important also for the explosion mechanism itself. The stellar rotation may be the simplest. Unsteady accretions of turbulent matter due to strong convection in the last stage of stellar evolution could be the seed perturbations for the asymmetry of CCSNe, which would set off different kinds of hydrodynamical instabilities during the stalled-shock phase and enhance neutrino heating and increase turbulent energies in the post-shock flow \citep{2015ApJ...808L..21C}. The nascent field of gravitational wave astronomy will be capable of directly investigating such asymmetric dynamics in the vicinity of proto-neutron star (PNS). The multi-dimensional modelling of CCSNe is hence indispensable to unveil the mechanism of CCSNe.


The neutrino radiation-hydrodynamic simulations of CCSNe have made a remarkable progress with ever increasing computational resources in the last few decades (for the current status, see e.g., \citet{2008ApJ...685.1069O,2014ApJ...786...83T,2015arXiv150106330K,2015ApJ...800...10D,2014arXiv1409.5779B,2015ApJ...807L..31L,2012ApJ...756...84M,2015arXiv150102999J}). Although the present interest of supernova society is directed mainly to 3D hydrodynamical aspects, neutrino transfer is certainly one of the most important ingredients in realistic modelling of CCSNe. Since neutrinos are not in thermal equilibrium with matter in general, the time evolution of the neutrino distribution function at each spatial location should be determined in principle by solving the Boltzmann equation in six-dimensional phase space with both special and general relativistic effects taken into account properly.

 Both high numerical cost and technical difficulties have prevented us from conducting the ab-initio simulations, however. As a matter of fact, various approximations have been employed even in the most sophisticated multi-dimensional simulations, which include the multi-group flux-limited diffusion (MGFLD) \citep{1998ApJ...495..911M,2005ApJ...626..317W,2013ApJS..204....7Z,2015ApJ...800...10D}, isotropic diffusion source approximation (IDSA) \citep{2009ApJ...698.1174L,2010PASJ...62L..49S,2014ApJ...786...83T,2016ApJ...817...72P}, two-moment method \citep{2015arXiv150102999J,2015arXiv151200113S,2015arXiv151107443O}, fast-multi-group transport (FMT) method \citep{2015MNRAS.448.2141M}, variable Eddington tensor method \citep{2012ApJ...756...84M} and 2D Boltzmann transport method without $v/c$ corrections \citep{2004ApJ...609..277L,2008ApJ...685.1069O}; some of them \citep{2012ApJ...756...84M,2014arXiv1409.5779B,2015ApJ...807L..31L} employ the ray-by-ray plus approximation further.
 It should be stressed that some of these approximations have been validated only under spherical symmetry and their performance in non-spherical situations, which no doubt prevail in the post-bounce supernova core, are still uncertain (see some also \citet{2006A&A...447.1049B,2015arXiv151200113S} for the validation of the ray-by-ray approximation). In fact, some recent simulations have yielded outcomes, that seem at odds with each other, the reason for which may be the different approximations they adopted for neutrino transfer \citep{2012ApJ...756...84M,2014ApJ...786...83T,2014arXiv1409.5779B,2015arXiv151200113S}. Multi-dimensional simulations with a Boltzmann solver, to which we refer in the following as Boltzmann-Hydro simulations, are hence indispensable to validate these approximations. They are obviously crucial to address the CCSNe mechanism.


Motivated by these facts, we have tackled the development of a Boltzmann-Hydro solver for the last few years. \citet{2012ApJS..199...17S} constructed a three dimensional Newtonian Boltzmann solver, which was latter coupled to a hydrodynamics solver with self-gravity and, more importantly, was extended to fully accommodate special relativity \citep{2014ApJS..214...16N}. The latter paper demonstrated in 1D simulations of CCSNe the capability of our new code based on two different energy grids: Lagrangian-remapping grid and laboratory-fixed grid. Having been fine-tuned for massive parallel supercomputers, this code in the 2D version is currently being run on the K supercomputer for axisymmetric simulations of CCSNe. This paper is based on our solution to the problem we encountered in these computations.

The problem is the following: the nascent PNS starts to receive random kicks shortly after core bounce when the matter that has experienced prompt convection falls onto the PNS. It is further kicked around later by the matter that has undergone hydrodynamical instabilities such as the standing accretion shock instability (SASI) or neutrino-driven convection (see also \citet{1994A&A...290..496J,2010PhRvD..82j3016N,2006A&A...457..963S}). The PNS then moves at velocities of the order of $100$km/s and is temporarily dislocated by several kilometers from the coordinate center (see also \citet{2012MNRAS.423.1805N}). It is also important to remember that if the shock wave is successfully revived, asymmetric ejecta will certainly produce PNS kicks. The fixed polar coordinates are not appropriate to follow these proper motions of PNS, since dislocated spheres cannot be reproduced very well on this grid. Indeed, we found that the simulation without any special treatment either crashed or resulted in unphysical outcomes once the PNS moves a few km away from the mesh center.

 The previous studies treated this problem rather pragmatically. For example, \citet{2013ApJ...770...66H,2014ApJ...786...83T,2014arXiv1409.5779B,2015ApJ...807L..31L,2015MNRAS.453..287M}) restricted the motion of PNS by artificially imposing spherical symmetry in the PNS. Although no numerical problem may have occured in these methods, it might have discarded some potentially important physics along. On the other hand, MPA group employed a moving-mesh technique, adding an extra velocity to the advection terms in the equations of motion that compensates the PNS motion \citep{2006A&A...457..963S}. In their experimental simulations \citet{2010PhRvD..82j3016N} also used a similar moving-mesh method to track the PNS motion, remapping the grid. It is stressed, however, that the latter two groups completely ignored the effects of the moving-mesh on neutrino transfer.


In this paper, we propose an entirely new method to precisely treat this issue not only for hydrodynamics but also for neutrino transfer. The basic idea is like in the previous papers \citep{2006A&A...457..963S,2010PhRvD..82j3016N} to globally translate the coordinates so that the mass center of PNS should stay always very close to the mesh center. It should work, since the PNS remains almost spherical even when it oscillates or runs away\footnote{This is because the recoil velocity is normally much lower than the sound velocity in PNS.}. The important thing is that the basic equations both on neutrino transfer and hydrodynamics should be modified on this moving grid, since it is not an inertial system. Hydrodynamics equations are easy to extend (see e.g., \citet{2006A&A...457..963S}) \footnote{In the Newtonian case, we add terms including the acceleration in the momentum and energy conservation equations. For GR cases, we do not need to modify the basic equations, since the acceleration has already been included through the shift vector. See the text for more details.}, while the modification of Boltzmann equation is more complicated because the coordinate acceleration affects the neutrino transport in non-trivial ways in the six-dimensional phase space. Then the general relativistic description is a natural and probably the unique choice for handling various effects correctly. We deal with this within the 3+1 formulation of GR, using the conservation form of the general relativistic Boltzmann equation \citep{2014PhRvD..89h4073S}. This means that the current upgrade of the neutrino transport module in our code is equivalent to a GR extension, which was actually planned as the next improvement to our Boltzmann solver. It should be mentioned, however, that in this paper the GR transport code is applied only to the flat Minknowski spacetime, since the treatment of gravity in our code is still Newtonian; the Newtonian version of the hydrodynamics code is employed in this work for the same reason (see also Section~\ref{sec:Feedback}). The GR capability of the code will be demonstrated elsewhere \citep{Nagainprep}.





This paper is organized as follows. In Section~\ref{sec:baseeq}, we reformulate the Boltzmann equation on the moving-mesh with the language of the 3+1 formalism of GR. We also explain the numerical implementation of this GR extension to our previous special relativistic (SR) code in Section~\ref{sec:extension}. Then, the feedbacks to hydrodynamics are described in Section~\ref{sec:Feedback}. We validate our new method with two tests: PNS oscillations around and runaways from its original position. The results are presented in Section~\ref{sec:twodsim}. Finally we conclude this paper with a summary in Section~\ref{sec:summary}. Throughout this paper, Greek and Latin subscripts denote space-time and space components, respectively. We use the metric signature of $- + + +$. Unless otherwise stated, we work in units with $c=G=1$, where $c$ and $G$ are the light speed and gravitational constant, respectively.

\begin{figure*}
\vspace{15mm}
\epsscale{1.0}
\plotone{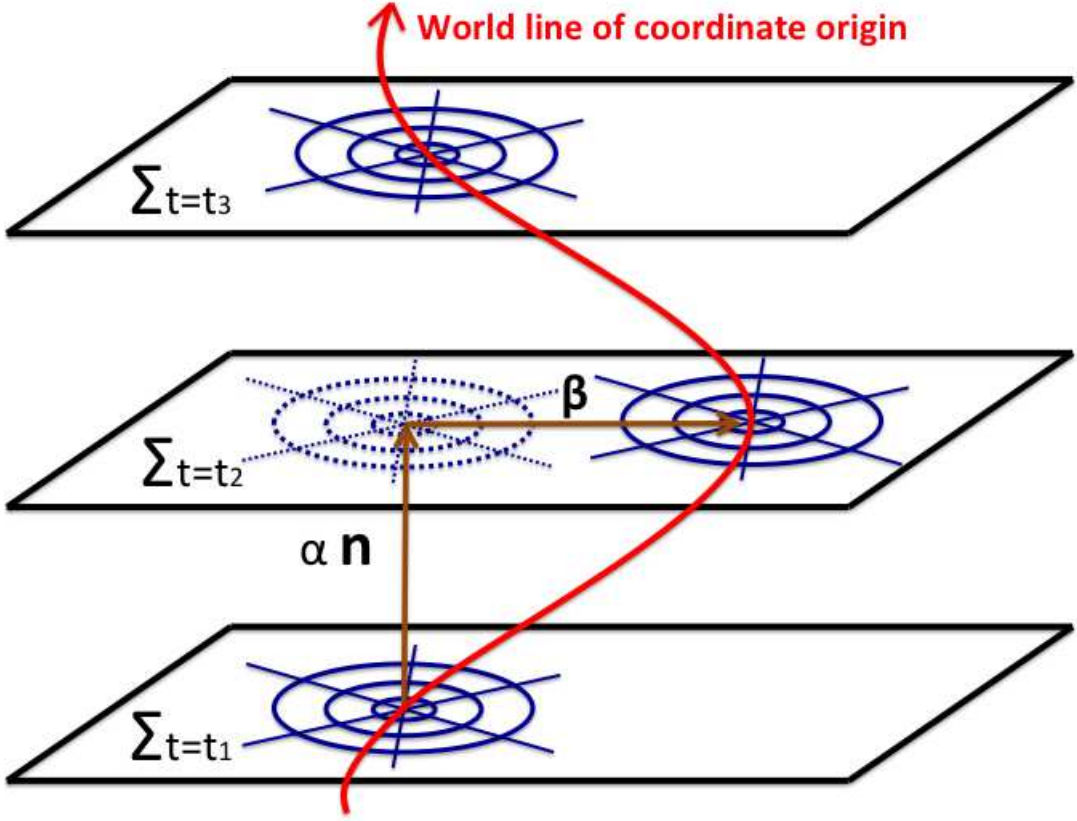}
\caption{Schematic picture for the moving-mesh in the 3+1 foliation of spacetime. Red line indicates the world line of coordinate origin. Concentric circles and radial rays on each spatial hypersurface ($\Sigma_{t}$) denote the polar grid. The coordinate origin traces the motion of PNS. See the text in more detail.
\label{fig:shift}} 
\end{figure*}

\section{Basic Equations}\label{sec:baseeq}
Following \citet{2014PhRvD..89h4073S}, we start with the conservation form of the Boltzmann equation in general relativity:
\begin{eqnarray}
&& \frac{1}{\sqrt{-g}} \left. \frac{\partial}{\partial x^{\alpha}} \right|_{q_{i}} 
\Biggl[  \Bigl( e^{\alpha}_{(0)} + \sum^{3}_{i=1} \ell_{i} e^{\alpha}_{i} \Bigr) \sqrt{-g} f   \Biggr] \nonumber \\
&& - \frac{1}{\nu^2} \frac{\partial}{\partial \nu}( \nu^3 f \omega_{(0)}  )
+ \frac{1}{{\rm sin}\bar{\theta}} \frac{\partial}{\partial \bar{\theta}} 
( {\rm sin}\bar{\theta} f \omega_{(\bar{\theta})} ) \nonumber \\
&& + \frac{1}{ {\rm sin}^2 \bar{\theta}} \frac{\partial}{\partial \bar{\phi}} (f \omega_{(\bar{\phi})}) = S_{\rm{rad}}, \label{eq:basicBoltz}
\end{eqnarray}
where $g, x^{\alpha}$ are the determinant of the metric, coordinates of spacetime, respectively, and $f$ is the neutrino distribution function; $e^{\alpha}_{(\mu)} (\mu = 0, 1, 2, 3)$ denote a set of the tetrad bases for a local orthonormal frame; $\ell_{i}$ are directional cosines for the direction of neutrino propagation with respect to $e^{\alpha}_{(i)}$ (see Fig. 1 in \citet{2014PhRvD..89h4073S}). The three components of $\ell_{i}$ can be written as
\begin{eqnarray}
&& \ell_{(1)} = {\rm cos} \hspace{0.5mm} \bar{\theta}, \nonumber \\
&& \ell_{(2)} = {\rm sin} \hspace{0.5mm} \bar{\theta}   {\rm cos} \hspace{0.5mm} \bar{\phi}, \nonumber \\
&& \ell_{(3)} = {\rm sin} \hspace{0.5mm} \bar{\theta}   {\rm sin} \hspace{0.5mm} \bar{\phi}, \label{eq:el}
\end{eqnarray}
where $\bar{\theta}$ and $\bar{\phi}$ stand for the polar and azimuthal angles \citep{1966AnPhy..37..487L}. We further define coordinates $q_{i}$ in momentum space- $q_{1}~=~\nu, q_{2}~=~\bar{\theta}$ and $q_{3}~=~\bar{\phi}$ with $\nu$ being the neutrino energy in this local orthonormal frame and also expressed as $\nu \equiv  - p_{\alpha} e^{\alpha}_{(0)}$ with the four momentum of neutrino, $p^{\alpha}$. In this paper neutrinos are assumed to be massless. $\omega_{(0)}, \omega_{(\bar{\theta})}, \omega_{(\bar{\phi})}$ are given as
\begin{eqnarray}
&& \omega_{(0)} \equiv \nu^{-2} p^{\alpha} p_{\beta} \nabla_{\alpha} e^{\beta}_{(0)}, \nonumber \\
&& \omega_{(\bar{\theta})} \equiv \sum^{3}_{i=1} \omega_{i} \frac{ \partial \ell_{(i)} }{\partial \bar{\theta} }, \nonumber \\
&& \omega_{(\bar{\phi})} \equiv \sum^{3}_{i=2} \omega_{i} \frac{ \partial \ell_{(i)} }{\partial \bar{\phi} }, \nonumber \\
&& \omega_{i} \equiv \nu^{-2} p^{\alpha} p_{\beta} \nabla_{\alpha} e^{\beta}_{(i)}. \label{eq:Omega}
\end{eqnarray}
 As shown in \citet{2014PhRvD..89h4073S}, these $\omega$'s can be expressed with the Ricci rotation coefficients. $S_{\rm{rad}}$ on the right hand side of Eq.~(\ref{eq:basicBoltz}) originates from the collision term for neutrino-matter interactions.

In the 3+1 formulation of GR, the line element is expressed as
\begin{eqnarray}
ds^2 = (- \alpha^2 + \beta^k \beta_k ) dt^2 + 2 \beta_i dt dx^i + \gamma_{ij} dx^i dx^j, \label{eq:lineeleme}
\end{eqnarray}
where $\alpha, \beta^{i}$ and $\gamma_{ij}$ denote the lapse function, shift vector and spatial 3-metric, respectively. In our extended Boltzmann code, the time-like basis $e^{\alpha}_{(0)}$ is chosen so that it should coincide with the unit vector $n^{\alpha}$ normal to the spatial hypersurface with $t={\rm const}$. This choice is a natural extension from our previous SR Boltzmann solver (see Section~\ref{sec:extension} for more details). Then three other spatial tetrad bases are taken so that they should be tangential to the spatial hypersurface. In this paper we assume that the spacetime is flat and is foliated with flat spatial hypersurfaces, on which we deploy the polar coordinates ($x^1 = r, x^2 = \theta, x^3 = \phi$). Then non-vanishing components of the 3 metric are $\gamma_{rr} = 1, \gamma_{\theta \theta} = r^{2}$ and $\gamma_{\phi \phi} = r^2 {\rm sin}{\theta}^2$. The spatial tetrad bases are chosen so that the $e_{(1)}$ be parallel to the radial coordinate, and $e_{(2)}$ be tangential to the surface spanned by $\partial_t$ and $\partial_{\theta}$, and $e_{(3)}$ be orthogonal to the other two:
\begin{eqnarray}
&& e^{\alpha}_{(1)} = (0, \gamma^{-1/2}_{rr}, 0, 0 ) \nonumber \\
&& e^{\alpha}_{(2)} = \Biggl(0, -\frac{\gamma^{-1/2}_{r \theta}}{\sqrt{\gamma_{rr} (\gamma_{rr} \gamma_{\theta \theta} - \gamma^2_{r \theta})}}, \sqrt{ \frac{\gamma_{rr}}{ \gamma_{rr} \gamma_{\theta \theta} - \gamma^2_{r \theta} } }, 0 \Biggr) \nonumber \\
&& e^{\alpha}_{(3)} = \Biggl(0, \frac{\gamma^{r \phi}}{\sqrt{\gamma^{\phi \phi}}} , \frac{\gamma^{\theta \phi}}{\sqrt{\gamma^{\phi \phi}}}, \sqrt{\gamma^{\phi \phi}} \Biggr). \label{eq:polartetrad}
\end{eqnarray}
 We refer to this orthonormal frame as the O-frame in the following. In accord with the above foliation of spacetime we set $\alpha = 1$. We utilize the shift vector to deal with the motion of the spatial coordinates (see Figure~\ref{fig:shift}). In fact, we set $\beta^{i} = \bar{V}^i$, where $\bar{V}^i$ is approximately the velocity of PNS measured in the O-frame (see the next section for details). Note that the employment of the globally uniform shift vector in this paper should be compatible with the use of other gauge conditions for the shift vector in possible applications of the current formulation to (dynamical) curved spacetimes. This completes the description of Eq.~(\ref{eq:basicBoltz}). We now turn to its numerical implementations.

\section{Numerical Implementation}\label{sec:extension}

\subsection{Shift vector} \label{subsec:shiftvector}
 Let us suppose that the basic equations are somehow finite-differenced and all hydrodynamics and space-time quantities are obtained up to the $n$-th time step. The average velocity of PNS at this time step ($V^{i(n)}$) is then given via the linear momentum (P) and mass of PNS (M) as
\begin{eqnarray}
&& V^{i(n)} = \frac{P^{i(n)}}{M^{(n)}}, \label{eq:PNSvelo} \nonumber \\
&& P^{i(n)} \equiv \int \rho^{(n)} v^{i(n)}_{o} dV_{\rm{PNS}}, \nonumber \\
&& M^{(n)}  \equiv \int \rho^{(n)} dV_{\rm{PNS}}, \label{eq:PNS_MomandMass}
\end{eqnarray}
where $\rho$, $v^{i}_{o}$ and $dV_{\rm{PNS}}$ denote the density, 3-velocity of matter (measured in the O-frame) and volume element in PNS, respectively. The PNS is defined to be the region, where the angle-averaged density ($\bar{\rho}$) is larger than $10^{13} {\rm g/cm}^3$. The time derivative of the velocity, or the acceleration of PNS, at the same time step is given by the following relation:
\begin{eqnarray}
\frac{dV^{i(n+\frac{1}{2})}}{dt} = \frac{( V^{i(n+1)} - V^{i(n)} )}{\Delta t^{(n)}}, \label{eq:accelPNS}
\end{eqnarray}
where $\Delta t^{(n)}$ is the interval between the $(n+1)$-th and $n$-th time steps.

Note that we do not use these $V^i$ and $dV^i/dt$ as they are for the following reasons. First, $V^{i}$ obtained in this way shows glitches from time to time when the PNS surface traverses an interface of the radial mesh points. Second, if the tracking of PNS motions were to be perfect, the acceleration of PNS should be determined iteratively, since the velocity of PNS at the next time step should be consistent with this acceleration but is obtained only after the advancement of the step. Such iterative process would be very time-consuming. Fortunately, however, it is unnecessary to exactly trace the motion of PNS and it turns out that the following approximate treatment suffices to deal with the proper motion of PNS.

We define the approximate PNS velocity $\bar{V}^{i}$ as follows:
\begin{eqnarray}
&&\bar{V}^{i(n+1)} = \bar{V}^{i(n)} + \frac{d\bar{V}^{i(n)}}{dt}  \Delta t^{(n)}, \label{eq:efPNSv}
\end{eqnarray}
with
\begin{eqnarray}
&&\frac{d\bar{V}^{i(n)}}{dt} = \frac{dV^{i(n-\frac{1}{2})}}{dt} + C^{(n)} + D^{(n)}, \nonumber \\
&& C^{(n)} \equiv ( V^{i(n)} - \bar{V}^{i(n)} )/ T, \nonumber \\
&& D^{(n)} \equiv X^{i(n)}_{m} / T^2, \label{eq:modacc}
\end{eqnarray}
$dV^{i(n-\frac{1}{2})}/dt$ is given by Eq.~(\ref{eq:accelPNS}) (but the backward difference); $C^{(n)}$ and $D^{(n)}$ are the terms that allow some deviations of the coordinate velocity and/or origin (denoted here by $X^{i(n)}_{m}$) from those of PNS and thus avoid the glitch; $T$ is the recovering time and is taken to be $0.1$ms in this paper. $C^{(n)}$ and $D^{(n)}$ also prevent secular drifts of PNS. In fact $C^n$ works as a damper to prohibit large differences between two velocities, whereas $D^n$ serves as an attractor to ensure that the coordinate origin tends to the mass center of the PNS. As an additional measure to ensure smooth coordinate motions, we do not update the value of $d\bar{V}^{i(n)}/dt$ when the PNS surface trespasses an interface of radial mesh points. As demonstrated later, we find that employing $\bar{V}$ as the shift vector in combination with the evaluation of $d\bar{V}^{i(n)}/dt$ given above is indeed sufficient to solve the problems with the proper motion of PNS.

Although the shift vector field thus obtained is spatially uniform, its derivatives with respect to $r, \theta$ and $\phi$ are non-vanishing, since the coordinates are curvilinear. The explicit form of the Ricci rotation coefficients are rather involved (although calculations are straightforwardly) numerically in the code. This will be useful indeed, since we are required to evaluate Ricci rotation coefficients for numerically obtained metrices in truly GR simulations.

\subsection{Modifications to SR code} \label{subsec:modificSR}

Although the GR Boltzmann equation, Eq.~(\ref{eq:basicBoltz}), has a simple form, the consistent treatment of the advection and collision terms is complicated even for the flat spacetime. In \citet{2014ApJS..214...16N}, we surmounted the difficulties by introducing two energy grids: {\it Lagrangian remapped grid} (LRG) and {\it Laboratory fixed grid} (LFG). We also devised for the SR code some other numerical techniques (e.g., a semi-implicit method for temporal sweep). It is therefore desirable in the GR extension to the current SR code that we should retain these features as much as possible. In the following, we describe how we achieved it.

We first consider the Boltzmann equation (\ref{eq:basicBoltz}) in flat spacetime. The tetrad bases, Eqs.~(\ref{eq:polartetrad}), are reduced in this case to
\begin{eqnarray}
&& e^{\alpha (F)}_{(0)} = (1, 0, 0, 0 ), \nonumber \\
&& e^{\alpha (F)}_{(1)} = (0, 1, 0, 0 ), \nonumber \\
&& e^{\alpha (F)}_{(2)} = \Biggl(0, 0, \frac{1}{r}, 0 \Biggr), \nonumber \\
&& e^{\alpha (F)}_{(3)} = \Biggl(0, 0, 0, \frac{1}{r {\rm sin} \theta } \Biggl), \label{eq:polartetrad_flat}
\end{eqnarray}
where the subscripts "F" hereafter implies quantities in the flat spacetime. Then we can evaluate the $\omega$ variables in Eq.~(\ref{eq:Omega}) as: 
\begin{eqnarray}
&& \omega_{(0)}^{(F)} = 0, \nonumber \\
&& \omega_{(\bar{\theta})}^{(F)} = - \frac{{\rm sin} \bar{\theta} }{r}, \nonumber \\
&& \omega_{(\bar{\phi})}^{(F)} = - \frac{ {\rm cot}\theta }{r} {\rm sin}^3 \bar{\theta} \hspace{0.5mm} {\rm sin} \bar{\phi}. \label{eq:Omega_flat}
\end{eqnarray}
Substituting these results into Eq.~(\ref{eq:basicBoltz}) and using the determinant of the metric for the fixed polar coordinates in the flat spacetime ($\sqrt{-g^{(F)}} = r^2 {\rm sin}\theta$), we reproduce the SR Boltzmann equation we employed in \citet{2014ApJS..214...16N}.

As we mentioned earlier, in the GR extension we want to retain the various features already implemented in our SR Boltzmann-Hydro code. This is most easily achieved by casting Eq.~(\ref{eq:basicBoltz}) into the following form:
\begin{eqnarray}
&& \frac{V}{\sqrt{-g^{(F)}}} \left. \frac{\partial}{\partial x^{\alpha}} \right|_{q_{i}} 
\Biggl[ K^{\alpha}   \biggl\{ \Bigl( e^{\alpha}_{(0)} + \sum^{3}_{i=1} \ell_{i} e^{\alpha}_{i} \Bigr) \sqrt{-g} \biggr\}^{(F)}  f   \Biggr] \nonumber \\
&& - \frac{1}{\nu^2} \frac{\partial}{\partial \nu}( \nu^3 f \omega_{(0)}  )
+ \frac{1}{{\rm sin}\bar{\theta}} \frac{\partial}{\partial \bar{\theta}} 
(   {\rm sin}\bar{\theta} f \omega_{(\bar{\theta})}^{(F)} 
 +  {\rm sin}\bar{\theta} f \Delta \omega_{(\bar{\theta})}    ) \nonumber \\
&& + \frac{1}{ {\rm sin}^2 \bar{\theta}} \frac{\partial}{\partial \bar{\phi}} (f \omega_{(\bar{\phi})}^{(F)}  +  f \Delta \omega_{(\bar{\phi})}  ) = S_{\rm{rad}}, \label{eq:basicBoltz_nume}
\end{eqnarray}
with
\begin{eqnarray}
&& V \equiv \frac{ \sqrt{-g^{(F)}}  }{\sqrt{-g}  }, \nonumber \\
&& K^{\alpha} \equiv \frac{ \biggl\{ \Bigl( e^{\alpha}_{(0)} + \sum^{3}_{i=1} \ell_{i} e^{\alpha}_{i} \Bigr) \sqrt{-g} \biggr\}  }{ \biggl\{ \Bigl( e^{\alpha}_{(0)} + \sum^{3}_{i=1} \ell_{i} e^{\alpha}_{i} \Bigr) \sqrt{-g} \biggr\}^{(F)}  }, \nonumber \\
&& \Delta \omega_{(\bar{\theta})} \equiv \omega_{(\bar{\theta})} - \omega_{(\bar{\theta})}^{(F)}, \nonumber \\
&& \Delta \omega_{(\bar{\phi})} \equiv \omega_{(\bar{\phi})} - \omega_{(\bar{\phi})}^{(F)}. \label{eq:correcterms}
\end{eqnarray}
It should be apparent that these four variables in Eq.~(\ref{eq:correcterms}) can be regarded as the GR corrections to the SR equation. This allows us to directly utilize our SR Boltzmann code in the GR extension.
 Although we employ the moving spherical coordinates in the Minkowski spacetime in this paper, the GR-extended code can accommodate any metric and gauge conditions, evaluating various GR terms numerically.
 Note also that, unlike other advection terms, the energy-derivative term represents gravitational redshift, a purely GR effect, which we calculate on the LRG \citep{2014ApJS..214...16N}\footnote{Other advection terms are calculated on LFG (see \citet{2014ApJS..214...16N} for more details). Note also that the energy derivative term disappears in the current study for the moving-mesh in the flat spacetime, since $\omega_{(0)}$ becomes trivially zero.}.

The treatment of the collision terms is also similar to that in the flat spacetime. Since the collision terms can be most easily calculated in the fluid-rest frame, we first evaluate them in this frame and then Lorentz-transform them to the laboratory frame, which is identical to the O-frame in the current formulation. This is done with the tetrads corresponding to these frames. We denote the tetrad bases of the fluid-rest frame as $\mbox{\boldmath $\hat{e}$}_{\hat{\mu}}$, which is expressed with $\mbox{\boldmath $e$}_{({\nu})}$ as
\begin{eqnarray}
\mbox{\boldmath $\hat{e}$}_{(\hat{\mu})} \equiv \Lambda_{(\hat{\mu})}^{\hspace{2mm}(\nu)}  \mbox{\boldmath $e$}_{({\nu})}, \label{eq:deffluidtetrad}
\end{eqnarray}
where $\Lambda$ stands for the Lorentz boost transformation. The components of $\Lambda$ are given as
\begin{eqnarray}
&&\Lambda_{(\hat{\mu})}^{\hspace{2mm}(\nu)} = ( \Lambda^{(\hat{\mu})}_{\hspace{2mm}(\nu)} )^{-1} , \\
&&\Lambda^{(\hat{\mu})}_{\hspace{2mm}(\nu)} = 
\begin{pmatrix}
\gamma & - \gamma v^{(i)} \\
- \gamma v^{(j)} & \hspace{3mm} I^{3} + \frac{\gamma^2}{1+\gamma} v^{(i)} v^{(j)}
\end{pmatrix}
,
\end{eqnarray}
where $I^{3}$ denotes the $3 \times 3$ identity matrix, $v^{(i)}$ and $\gamma$ are defined with the tetrad bases of the laboratory frame and the fluid 4-velocity $\mbox{\boldmath $u$}$ as
\begin{eqnarray}
&&u_{(\mu)} \equiv \mbox{\boldmath $u$} \cdot  \mbox{\boldmath $e$}_{(\mu)}, \label{eq:4velotetdn} \\
&&u^{(\mu)} = \eta^{ \mu \nu } u_{(\nu)}, \label{eq:4velotetup} \\
&&v^{(i)} \equiv \frac{u^{(i)}}{u^{(0)}} , \label{eq:3velotetup} \\
&&\gamma \equiv u^{(0)}, \label{eq:deflorentzfac}
\end{eqnarray}
where $\eta^{\mu \nu}$ denotes the Minkowski metric. The 4-momentum of neutrino is also projected on $\mbox{\boldmath $\hat{e}$}_{(\hat{\mu})} $. Then the energy shift and aberration are determined by the Doppler factor given above and our SR formulation can be directly passed over to the GR-extended code (see Section 4 in \citet{2014ApJS..214...16N}).

\section{Feedback to Matter}\label{sec:Feedback}
In this section, we describe the feedback from neutrino interactions to hydrodynamics in detail. We first present the fully GR formulation, and then take the Newtonian limit. It is noted that the hydrodynamics part of our code is also fully GR but its Newtonian version is employed in this paper, since the self-gravity part of the code is still Newtonian.

 The basic equations for matter dynamics consist of the conservation laws of baryon number, energy-momentum and electron number, which are written as, respectively,
\begin{eqnarray}
(\rho_0 u^{\nu})_{;\nu} &=& 0, \label{eq:continuityeq} \\ 
T_{\rm{(hd)} \hspace{1.5mm} ;\nu}^{\mu \nu} &=& - G^{\mu}, \label{eq:TandGfinal} \\
N_{\rm{(e)} \hspace{1mm} ;\nu}^{\nu} &=& - \Gamma, \label{eq:NandGammafinal}
\end{eqnarray}
where $\rho_0$, $T_{\rm{(hd)}}^{\mu \nu}$ and $N_{(e)}^{\nu}$ denote the rest-mass density of baryons, energy-momentum tensor of matter and the electron number 4-current, respectively. The right-hand sides of the latter two equations represent the feedback, which are related to the collision term of Boltzmann equation, Eqs~(\ref{eq:basicBoltz}) or (\ref{eq:basicBoltz_nume}), as follows:
\begin{eqnarray}
G^{\mu} &\equiv& \sum_{\rm{i}} G_{\rm{i}}^{\mu}, \label{eq:Gsumdef} \\
G_{\rm{i}}^{\mu} &\equiv& \int p_{\rm{i}}^{\mu} \nu S_{\rm{rad} (\rm{i})} dV_p, \label{eq:Gdef} \\
\Gamma &\equiv& \Gamma_{\nu_{e}} - \Gamma_{\bar{\nu_{e}}}, \label{eq:Gammasumdef} \\
\Gamma_{\rm{i}} &\equiv& \int \nu S_{\rm{rad} (\rm{i})}  dV_p. \label{eq:Gammadef}
\end{eqnarray}
In these expressions, $dV_{p} (= \nu {\rm sin}\bar{\theta} d \nu d \bar{\theta} d \bar{\phi} )$ denotes the invariant volume in the momentum space. The subscript "$\rm{i}$" indicates the neutrino species, which we omit hereafter for simplicity.

In actual simulations, we first evaluate the tetrad components of $\mbox{\boldmath $G$}$ in the fluid-rest frame as
\begin{eqnarray}
\hat{G}_{(\hat{\mu})} &\equiv& \int \hat{p}_{(\hat{\mu})} \hat{\nu} \hat{S}_{\rm{rad}} dV_p, \label{eq:Gfluidrest}
\end{eqnarray}
where the hat indicates variables in the fluid-rest frame, i.e., $\hat{G}_{(\hat{\mu})} = \mbox{\boldmath $G$}  \cdot  \mbox{\boldmath $\hat{e}$}_{(\hat{\mu})}$ etc..
Then, the coordinate components of $\mbox{\boldmath $G$}$ can be expressed via $ \hat{G}_{(\hat{\nu})}$ and $\mbox{\boldmath $\hat{e}$}_{(\hat{\nu})}$ as:
\begin{eqnarray}
G^{\mu} = \sum_{\hat{\nu}} \hat{G}_{(\hat{\nu})} \hat{e}_{(\hat{\nu})}^{\mu}, \label{eq:coordinatecompoAup}
\end{eqnarray}
where $\hat{e}_{(\hat{\nu})}^{\mu}$ denotes the coordinate components of $\mbox{\boldmath $\hat{e}$}_{(\hat{\nu})}$.

In the 3+1 formalism the basic equations for matter dynamics can be expressed as
\begin{eqnarray}
&& \partial_t {\rho}_{\ast} + \partial_j\left( {\rho}_{\ast} v^j \right) = 0 ,
 \label{eq:conti3pra1} \\
&& \partial_t S_i + \partial_j\left( \alpha \sqrt{\gamma} \, T^{j}_{\rm{(hd)} \hspace{1mm} i}\right) 
 = \frac{1}{2} \alpha \sqrt{\gamma} \, T^{\alpha \beta} g_{\alpha \beta ,i} - G_i,
 \label{eq:Mon3pra1} \\
&& \partial_t {\tau}
 + \partial_i \left( {\alpha}^2 \sqrt{\gamma} \, T^{0i} 
                - {\rho}_{\ast} v^i \right) = s - \alpha^2 \sqrt{\gamma} G^{0},
 \label{eq:Ene3pra1} \\
&& \partial_t ({\rho}_{\ast} Y_{\rm e}) + \partial_j\left( {\rho}_{\ast} Y_{\rm e} v^j \right) = - \alpha \sqrt{\gamma} \hspace{0.5mm} \Gamma ,
 \label{eq:Lepcon3pra1}
\end{eqnarray}
 where various variables are defined as follows:
\begin{eqnarray}
 v^j & \equiv & \frac{u^j}{u^t} , \label{eq:threevelodef} \\
 {\rho}_{\ast} & \equiv & \alpha \sqrt{\gamma} \, \rho_0 u^t ,
 \label{eq:con1def} \\
 S_j & \equiv & \alpha  \sqrt{\gamma} \, T^0 \! _j 
 = {\rho}_{\ast} h u_j ,
 \label{eq:con2def} \\
 \tau & \equiv & \alpha^2  \sqrt{\gamma} \, T^{00} - {\rho}_{\ast}
 = {\rho}_{\ast} \alpha h u^t -  \sqrt{\gamma} \, p - {\rho}_{\ast} ,
 \label{eq:con3def} \\
 s & \equiv & \alpha \sqrt{\gamma} \,
 \biggl\{ \left( T^{00} \beta^i \beta^j + 2 T^{0i} \beta^j + T^{ij} \right) K_{ij} \nonumber \\
  && \hspace{5mm}  -\left( T^{00} \beta^i + T^{0i} \right) \partial_{i} \alpha \biggr\}, \label{eq:defsourceMom}
\end{eqnarray}
(see also Eq.~A2 in \citet{2008ApJ...689..391N}).
 In the above equations,
 $Y_{\rm e}$, $p$, $h$, $g_{\mu \nu}$ $\gamma$, $K_{ij}$ are electron fraction, pressure and specific enthalpy of matter, 4-dimensional metric of spacetime, the determinant of 3-dimensional metric of space and extrinsic curvature, respectively. 

In this paper, instead of employing these fully GR equations, we adopt their Newtonian approximations, which can be derived by taking the weak gravitational field limit, ignoring the time derivative of gravitational potential and the space derivatives of 3-dimensional space metric (see \citet{2011ApJ...731...80N}). Then basic equations can be reduced the spherical coordinates to
\begin{eqnarray}
\partial_{t}{\mbox{\boldmath $Q$}} + \partial_{j}{\mbox{\boldmath $U^{j}$}} = \mbox{\boldmath $W_{h}$} + \mbox{\boldmath $W_{i}$} + \mbox{\boldmath $W_{a}$}, \label{eq:hydroConservativeform_ac}
\end{eqnarray}
where each term is given as
\begin{eqnarray}
\hspace{0mm} \mbox{\boldmath $Q$} =
\left(
\begin{array}{c}
\sqrt{g} \rho \\
\sqrt{g} \rho v_{r} \\
\sqrt{g} \rho v_{\theta} \\
\sqrt{g} \rho v_{\phi} \\
\sqrt{g} ( e + \frac{1}{2} \rho v^2) \\
\sqrt{g} \rho Y_{e}
\end{array}
\right),
\end{eqnarray}
\begin{eqnarray}
\hspace{0mm} \mbox{\boldmath $U^{j}$} =
\left(
\begin{array}{c}
\sqrt{g} \rho v^{j} \\
\sqrt{g} (\rho v_{r} v^{j} + p \delta_r^{j})\\
\sqrt{g} (\rho v_{\theta} v^{j} + p \delta_{\theta}^{j})\\
\sqrt{g} (\rho v_{\phi} v^{j} + p \delta_{\phi}^{j})\\
\sqrt{g} ( e + p + \frac{1}{2} \rho v^2) v^{j} \\
\sqrt{g} \rho Y_{e} v^{j}
\end{array}
\right),
\end{eqnarray}
\begin{eqnarray}
\mbox{\boldmath $W_{h}$} =
\left(
\begin{array}{c}
0 \\
\sqrt{g} \rho \Bigl( - \psi_{,r} + r (v^{\theta})^2 + r {\rm sin}^2\theta (v^{\phi})^2 + \dfrac{2p}{r \rho} \Bigr)\\
\sqrt{g} \rho \Bigl( - \psi_{,\theta} r^2 + {\rm sin}\theta {\rm cos}\theta (v^{\phi})^2 + \dfrac{p  {\rm cos}\theta }{ \rho {\rm sin}\theta }  \Bigr)\\
- \sqrt{g} \rho \psi_{,\phi} \\
- \sqrt{g} \rho v^{j} \psi_{,j} \\
0
\end{array}
\right) \label{eq:Wh},
\end{eqnarray}
\begin{eqnarray}
\hspace{0mm} \mbox{\boldmath $W_i$} =
\left(
\begin{array}{c}
0 \\
- \sqrt{g} G^{r} \\
- \sqrt{g} G^{\theta} \\
- \sqrt{g} G^{\phi} \\
- \sqrt{g} G^{t} \\
- \sqrt{g} \Gamma
\end{array}
\right), 
\end{eqnarray}

\begin{eqnarray}
\hspace{0mm} \mbox{\boldmath $W_a$} =
\left(
\begin{array}{c}
0 \\
\sqrt{g} \rho \dot{\beta}_{r} \\
\sqrt{g} \rho \dot{\beta}_{\theta} \\
\sqrt{g} \rho \dot{\beta}_{\phi} \\
\sqrt{g} \rho v^{j} \dot{\beta}_{j} \\
0
\end{array}
\right),
\end{eqnarray}
(see also Eqs.(12)-(16) in \citet{2014ApJS..214...16N}).
In the above expressions, $\sqrt{g}(=r^2{\rm sin}\theta)$, $\psi$ and $\dot{\beta}_{j}$ denote the volume factor for the spherical coordinates, the Newtonian gravitational potential, and the time derivative of shift vector, respectively. Note that $\mbox{\boldmath $W_a$}$ represents the acceleration of the coordinates.

\section{Validation}\label{sec:twodsim}

\subsection{Numerical setup and input physics}\label{subsec:numericalsetup}

\begin{figure*}
\vspace{15mm}
\epsscale{1.0}
\plottwo{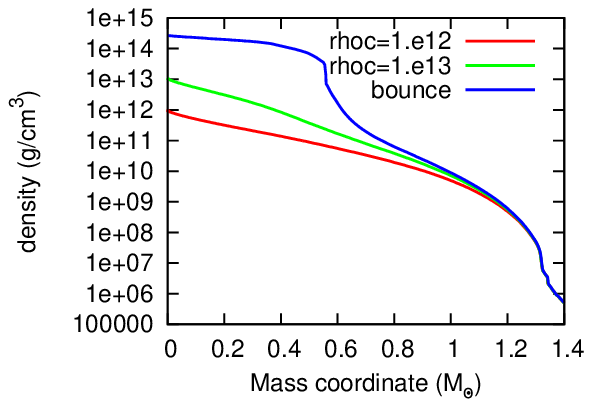}{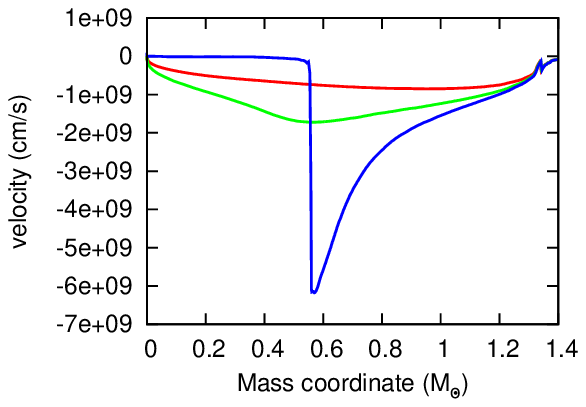}
\plottwo{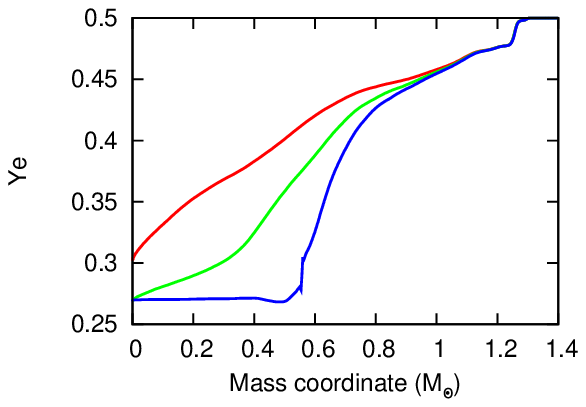}{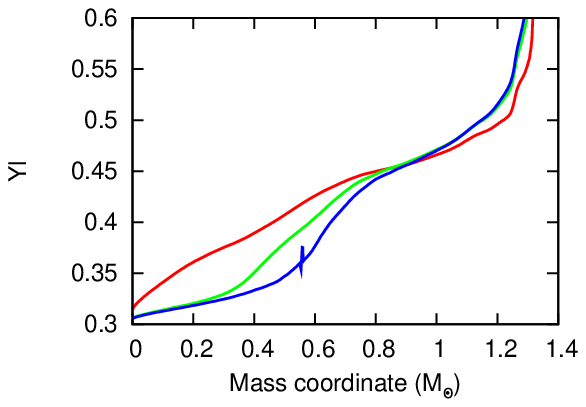}
\plottwo{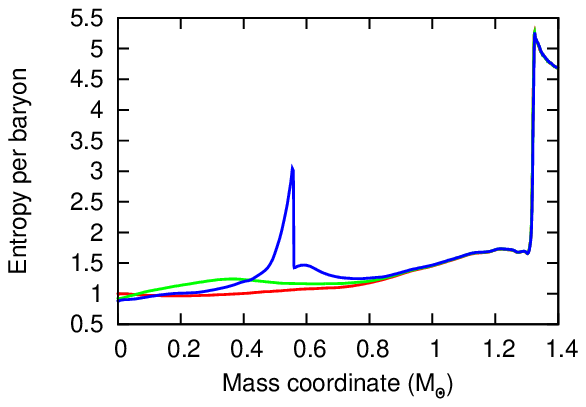}{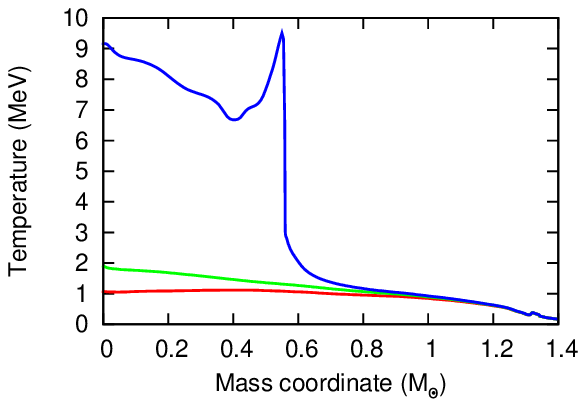}
\caption{Distributions of several hydrodynamical quantities in the pre-bounce phase obtained by the 1D spherically symmetric simulations. The horizontal axis is the mass coordinate.
\label{fig:1Dpreb}} 
\end{figure*}

\begin{figure*}
\vspace{15mm}
\epsscale{1.0}
\plottwo{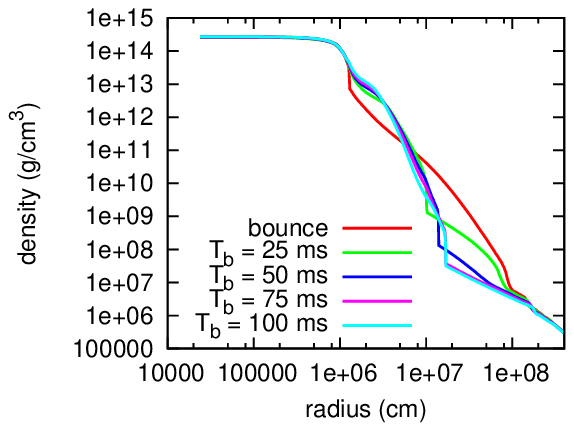}{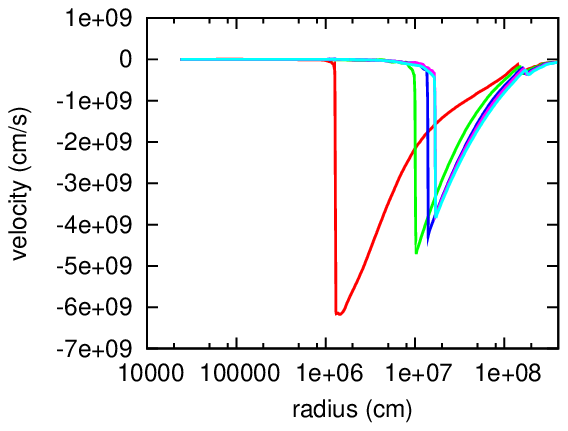}
\plottwo{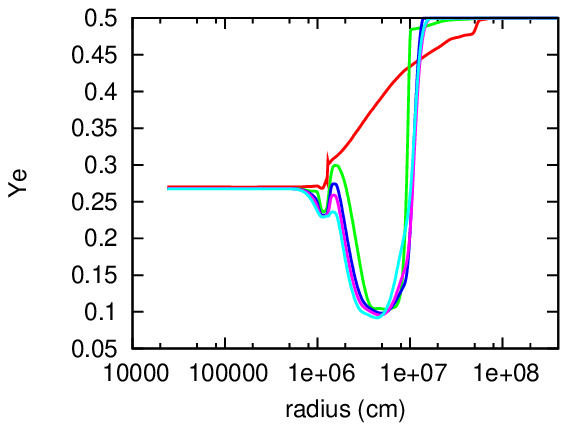}{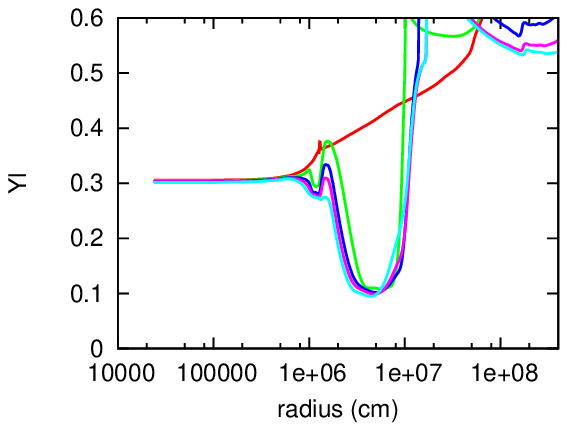}
\plottwo{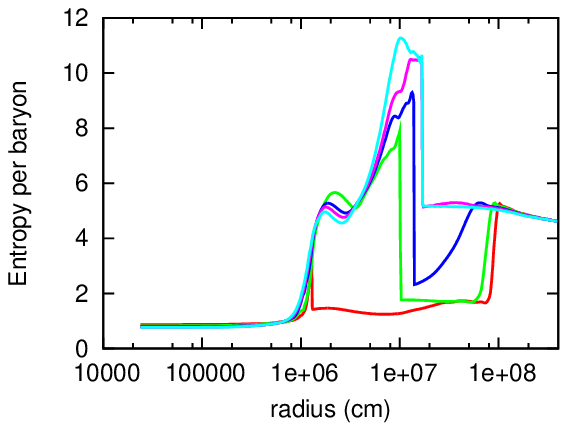}{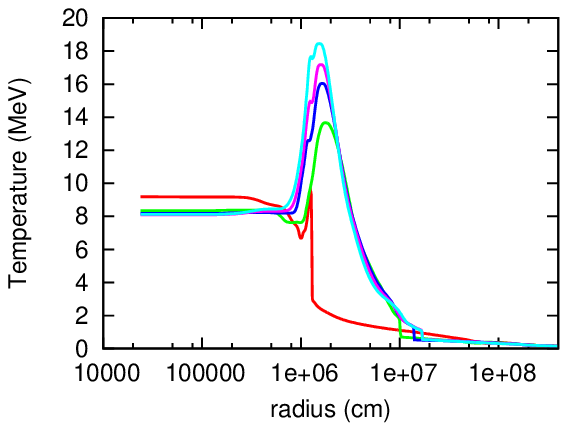}
\caption{Same as Figure~\ref{fig:1Dpreb} but for the post-bounce phase. Note that the horizontal axis is not the mass coordinate but the radius in these panels.
\label{fig:1Dpostb}} 
\end{figure*}

\begin{figure}
\vspace{15mm}
\epsscale{1.2}
\plotone{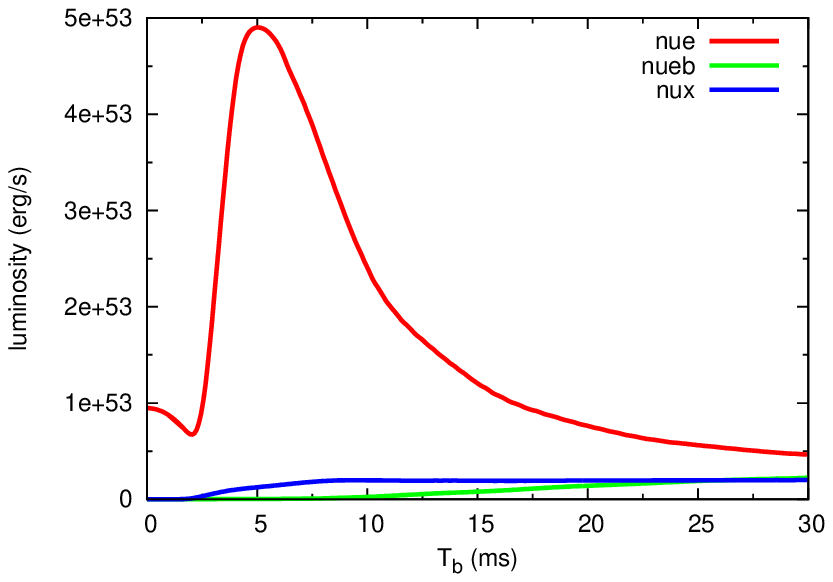}
\plotone{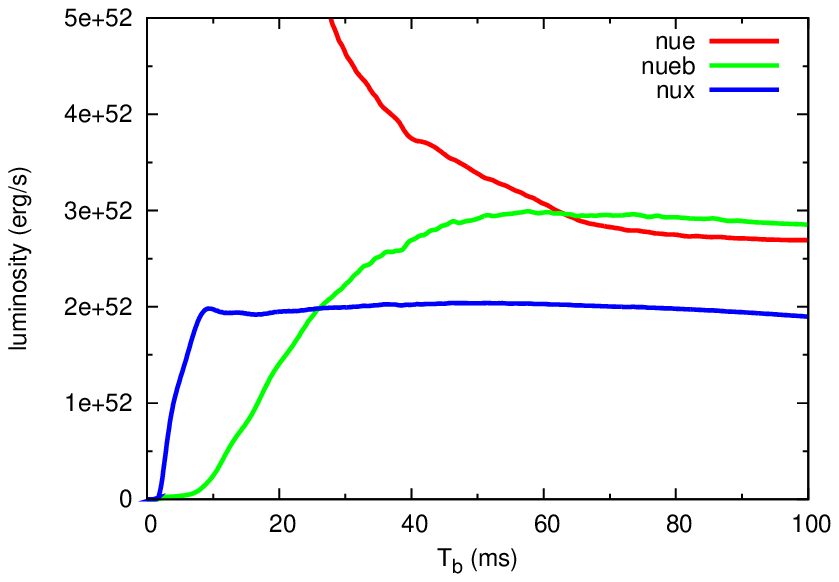}
\caption{Time evolutions of neutrino luminosity measured in the laboratory frame at $r=422$km. The red lines correspond to the electron-type neutrino whereas the green and blue lines are, respectively, for the electron-type anti-neutrino and heavy-lepton neutrinos. The upper panel zooms into the initial phase up to ${\rm{T}_{\rm{b}}}=30$ms, while the lower panel shows the longer evolutions until ${\rm{T}_{\rm{b}}}=100$ms in a different vertical scale.
\label{fig:lightcurve}} 
\end{figure}

\begin{figure}
\vspace{15mm}
\epsscale{1.2}
\plotone{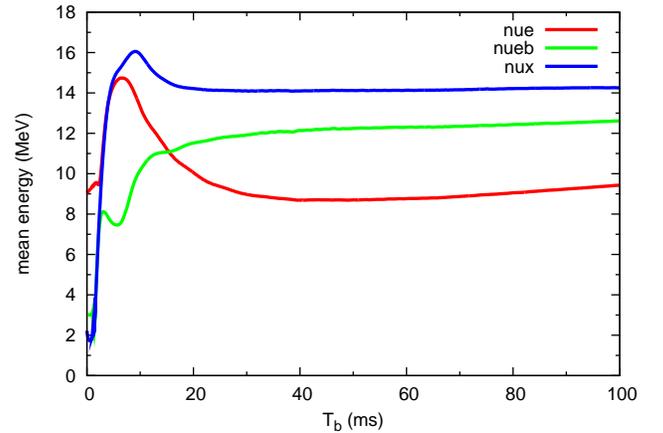}
\caption{Time evolutions of the mean energies measured in the laboratory frame at $r=422$km. Colors correspond to different neutrino species in as Figure~\ref{fig:lightcurve}.
\label{fig:aveene}} 
\end{figure}

\begin{figure}
\vspace{15mm}
\epsscale{1.2}
\plotone{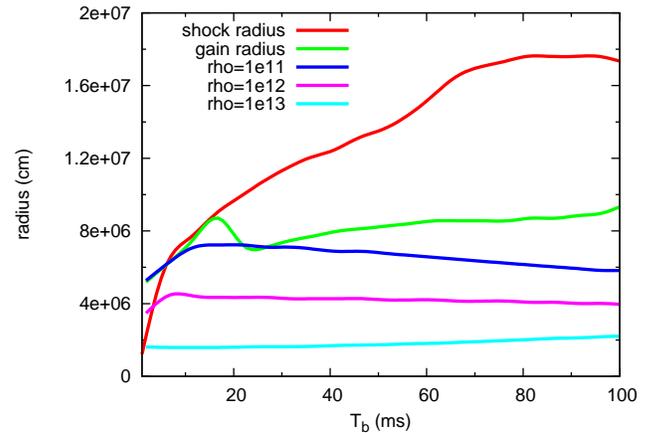}
\caption{Trajectories some of important radii in the post-bounce phase. The red line shows the shock position; the green line gives the gain radius; the other three lines present the trajectories of the points with the densities given in the figure.
\label{fig:variousradi}} 
\end{figure}

\begin{figure*}
\vspace{15mm}
\epsscale{1.0}
\plotone{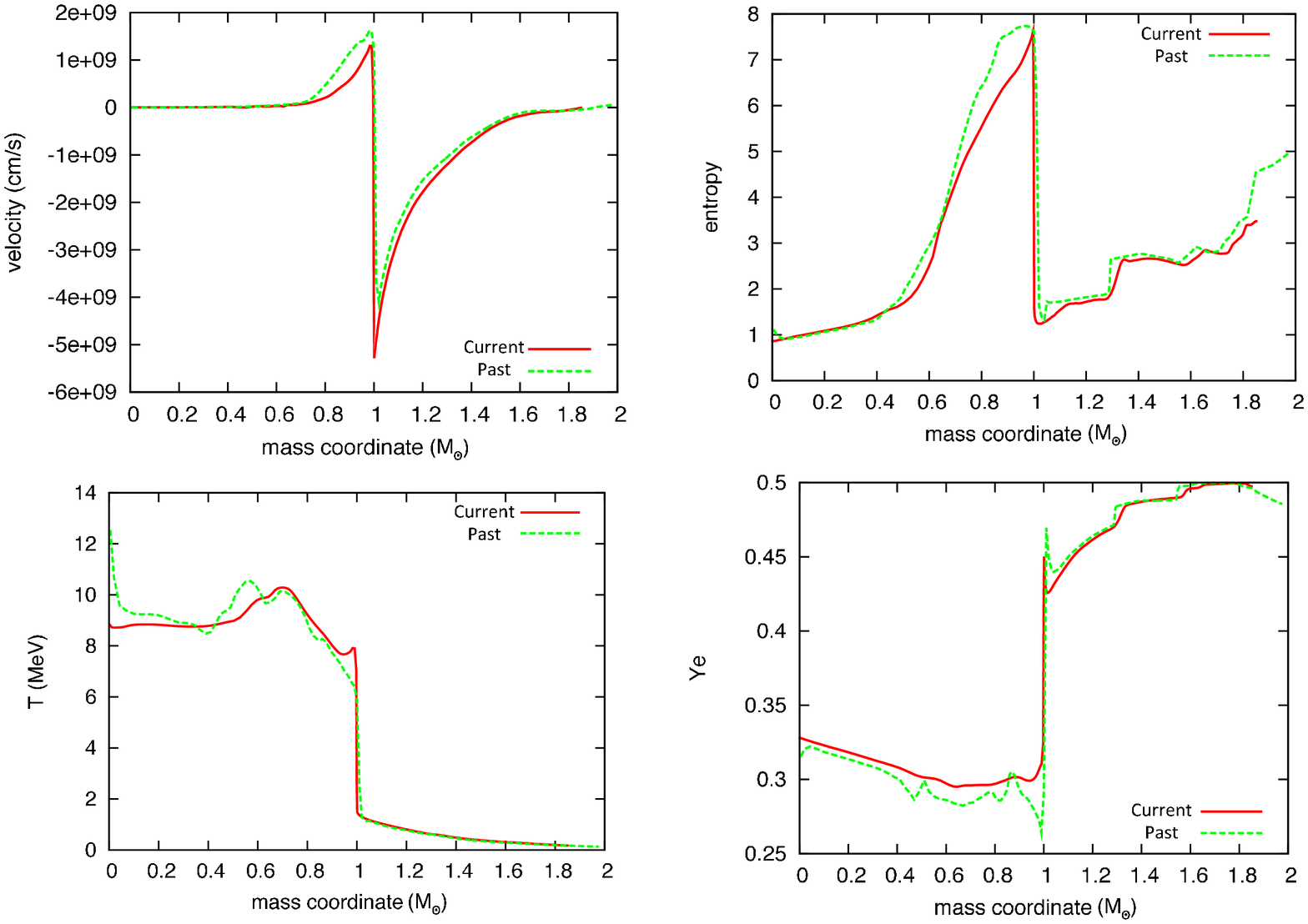}
\caption{Comparisons between our code and the one by \citet{2005ApJ...629..922S}
 in 1D simulations for the $15 M_{\sun}$ progenitor model by \citet{2002RvMP...74.1015W}. The radial velocity, entropy, temperature and electron fraction at core bounce are shown clockwise from the upper left panel. The red line corresponds to the results of the current code, while the green line represent ones for the other code.
\label{fig:compariSumi}} 
\end{figure*}

In this section, we validate our new formulation of the moving-mesh technique by performing 2D axisymmetric Boltzmann-Hydro simulations for the toy model of PNS oscillations and runaways from the original position. A thorough investigation of the code performance, including the GR capability, will be reported in the forthcoming paper \citep{Nagainprep}. As the initial conditions of these tests, we employ a configuration of a supernova core at 100ms after core bounce, which is obtained by a 1D spherically symmetric simulation of core-collapse, bounce and stagnation of shock wave with the same code. We map the resultant 1D configuration onto the 2D grid.

 We use a 11.2$\rm{M}_{\sun}$ progenitor \citep{2002RvMP...74.1015W}. For the 1D simulation, a non-uniform grid of $N_{r}=384$ points in the radial direction covers the region of $0 \leq r \leq 5000 $km, while the momentum space is discritized with a grid of $N_{\nu}=20$ points in the energy region of $0 \leq \nu \leq 300$ MeV and $N_{\bar{\theta}} = 10$ points covering the polar angle from 0 to $\pi$. For the 2D simulations, on the other hand, we deploy $N_{r}=192$ radial grid points and uniformly distributed $N_{\theta} = 32$ angular grid points in the entire meridian section; for the momentum space we use $N_{\nu}=20$ energy grid points and $N_{\bar{\theta}} (= 4) \times N_{\bar{\phi}} (= 4)$ angular grid points. Although this is admittedly a rather coarse mesh both in space and momentum space, it is not a serious issue for the current purpose: a proof-of-principle kind of code tests. We take into account three species of neutrinos: $\nu_e, \nu_{\bar{e}}$, and $\nu_{x}$, which are electron-type neutrinos, electron-type anti-neutrinos and heavy-lepton neutrinos ($\mu$ and $\tau$ neutrinos and their anti-particles collectively), respectively.

As for the input physics, our current Boltzmann-Hydro code has updated some treatments of microphysics from \citet{2012ApJS..199...17S,2014ApJS..214...16N}. One of them is an incorporation of the multi-nuclear species equation of state (EOS) by \citet{2011ApJ...738..178F,2013ApJ...772...95F}. This tabulated EOS provides us with not only thermodynamics quantities but also abundances of nuclei with mass numbers up to $A \sim 1000$ in nuclear statistical equilibrium or NSE, which are then employed to obtain the rate of electron captures by heavy nuclei (see below). Incidentally the EOS includes also the information on the abundances of light elements. We are currently studying possible roles of their interactions with neutrinos in the post-bounce phase of CCSNe.

Neutrino-matter interactions have been also improved from \citet{2012ApJS..199...17S}. One of the upgrades is the full account of non-isoenergetic scatterings between neutrinos and electrons and positrons. Unlike the spherically symmetric case, the interaction rate depends on $\bar{\phi}$, the fact which prevents a direct application of the method used in \citet{1993ApJ...410..740M,2005ApJ...629..922S}. We hence obtain the interaction rate by direct numerical integrations combined with the Chebychev expansions of Polylogarithms \citep{1970Kolbig}. It is important to note that an implicit treatment of non-isoenergetic scatterings is highly expensive both in computational time and memory. We hence handle the neutrino scatterings on electrons and positrons explicitly. Since they are rather minor, having smaller reaction rates, compared with other reactions such as emissions, absorptions and scatterings on nucleons, this poses no problem.

 Another important upgrade is the treatment of electron captures by nuclei as mentioned earlier. We tabulate the reaction rates based on the results by \citet{2010NuPhA.848..454J} and the approximation formula of \citet{2000NuPhA.673..481L} and \citet{2003PhRvL..90x1102L} with the mass fractions of heavy nuclei being taken from the Furusawa's EOS.

 As shown in \citet{2012ApJ...747...73L}, these two updates are critically important for CCSNe. As a matter of fact, the deleptonization would be erroneously suppressed during the infall phase if they were neglected, which would then result in a larger mass of the inner core and a stronger shock wave (see below and also \citet{1993ApJ...410..740M,2003PhRvL..90x1102L,2003PhRvL..91t1102H,2012ApJ...747...73L}).

 Figures~\ref{fig:1Dpreb}--\ref{fig:variousradi} display the result of the 1D simulation. Figure~\ref{fig:1Dpreb} plots the distributions of density, radial velocity, electron fraction, lepton fraction, entropy per baryon and temperature at different times in the pre-bounce phase. We find that the mass of the inner core is somewhat less than 0.6 $\rm{M}_{\sun}$ due to the unsuppressed deleptonization, which is consistent with other 1D computations (see, e.g., \citet{2012ApJ...747...73L,2015arXiv150807348S}). The post-bounce counter parts are shown in Figure~\ref{fig:1Dpostb} for different times, where ${\rm{T}_{\rm{b}}}$ denotes the time after bounce in this figure.

 Figure~\ref{fig:lightcurve} presents the energy fluxes measured in the laboratory frame (i.e., neutrino luminosities) at $r~=~422$~km. Note that the well-known bounce feature, i.e., a slight decrease followed by a quick rise in the luminosity of electron-type neutrino reaches this radius at ${\rm{T}_{\rm{b}}} \sim 4$ms. The prominent neutronization burst of electron-type neutrinos can be clear seen in the upper panel, while the luminosities of other neutrinos start to rise somewhat later. Note that the production of electron-type anti-neutrinos is initially suppressed by high electron fractions around the neutrino sphere ($Y_e \sim 0.3$). In fact, the luminosity of heavy-lepton neutrino goes up in $\sim~10$ms, while that of electron-type anti-neutrinos increases gradually for $\sim~50$ms. Figure~\ref{fig:aveene} shows the time evolution of the mean energy for each neutrino species at $r~=~422$~km. The mean energy is defined as
\begin{eqnarray}
E_{\rm{mean}} = \frac{\int f \nu^3 d \Omega d \nu}
                 {\int f \nu^2 d \Omega d \nu}. \label{eq:meanene}
\end{eqnarray}

We also show the trajectories of some important radii for the post-bounce phase in Figure~\ref{fig:variousradi}. The shock expands initially but is stagnated around $170$km at ${\rm{T}_{\rm{b}}} \sim 80$ms, while the gain radius starts to deviate from the shock radius at ${\rm{T}_{\rm{b}}} \sim 20$ms. The trajectories of the points that have the densities of $\rho = 10^{11}, 10^{12}$ and $10^{13}\rm{g/cm}^3$ are also shown in this figure. They will serve as a rough guide to the size of PNS as a function of time. All of these features are in qualitative agreement with those observed in previous studies (see e.g., \citet{2012ApJ...747...73L,2014ApJ...788...82M}). The data at ${\rm{T}_{\rm{b}}}=100$ms are used as the initial condition for the subsequent 2D simulations.

As a quick validation of the core part of our code, we show in Figure~\ref{fig:compariSumi} some results of the comparison with another code, in which we ran 1D simulations twice for the $15 M_{\sun}$ progenitor model by \citet{2002RvMP...74.1015W}, using the current code and another 1D but fully GR Boltzmann-Hydro code developed by \citep{2005ApJ...629..922S}. All input physics are identical between the two computations but the latter is fully general relativistic and employs the Lagrangian formulation. In the figure, we show some key quantities at core bounce. In each panel, the red (green) line gives the result of our new (\citet{2005ApJ...629..922S}) code. Considering the differences just mentioned, we think the two results agree with each other reasonably well.

\subsection{PNS Oscillation}\label{subsec:oscillation}

It is known that the center of PNS oscillates with velocities of the order of 100 km/s and periods of several tens of milliseconds (see e.g., \citet{2012MNRAS.423.1805N}). Mimicking such a situation, we start a 2D simulation by adding a velocity in the z-direction as $\Delta v_z=100$km/s in the region of $r < 30$km. This simulation is carried out for 100ms, which is long enough for the purpose of this study.

\begin{figure}
\vspace{15mm}
\epsscale{1.0}
\plotone{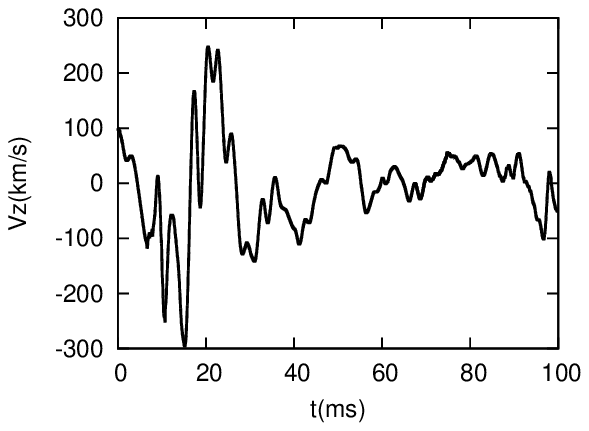}
\plotone{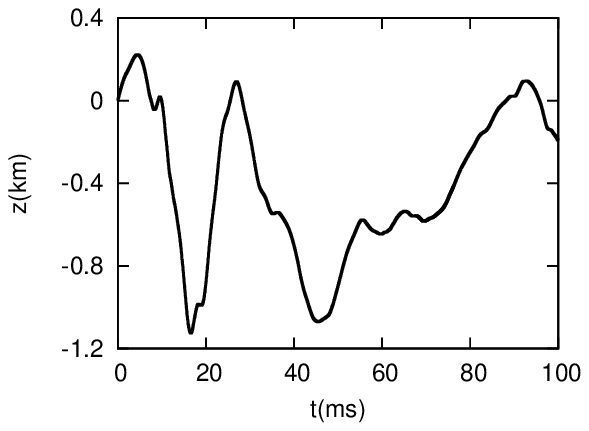}
\plotone{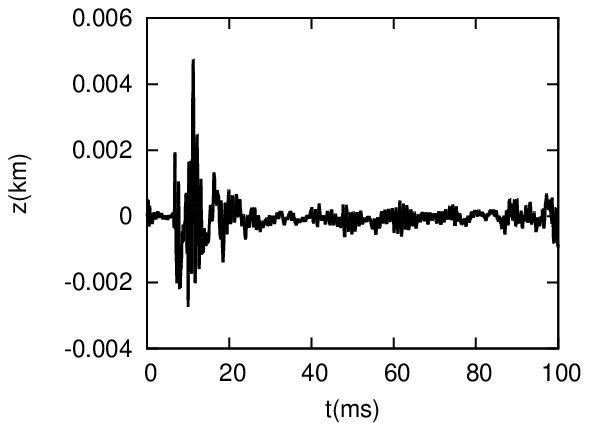}
\caption{Upper: Time evolution of PNS velocity. Middle: The trajectory of the center of moving-mesh. Lower: The spatial difference between the center of moving-mesh and the mass center of PNS.
\label{fig:Oscivelo}} 
\end{figure}

Figure~\ref{fig:Oscivelo} shows the time evolution of PNS velocity (upper panel), trajectory of the origin of moving-mesh (middle panel), and the z-coordinate of the mass center of PNS on this moving-mesh (lower panel). One can see from the middle panel that the PNS moves in the positive z direction initially owing to the velocity added at the beginning. It is also clear from the upper panel that the PNS is decelerated and the direction of motion is reversed after a few ms. It is noted the PNS moves from the original position by $\sim 1$km by the time of $t \sim 15$ms as shown in the middle panel. This is not a small distance and the subsequent evolution can not be calculated without the moving-mesh technique. As a matter of fact, we conducted the same simulation on the ordinary non-moving grid, and found that it ended up with a numerical crash with unphysical matter distributions around the coordinate origin.

 At $t \sim 20$ms the PNS again changes the direction of motion and returns to the origin. Although the period is variable in time, the PNS experiences two cycles of oscillations in this simulation. It is also important to note that the moving-mesh nicely traces the motion of PNS (see the lower panel). This leads to the successful Boltzmann-Hydro simulation on the spherical polar grid. We confirmed indeed that all the unphysical features observed in the simulation on the fixed mesh have gone with the moving-mesh technique.

\subsection{Runaway motions of PNS}\label{subsec:Kick}

\begin{figure}
\vspace{15mm}
\epsscale{1.0}
\plotone{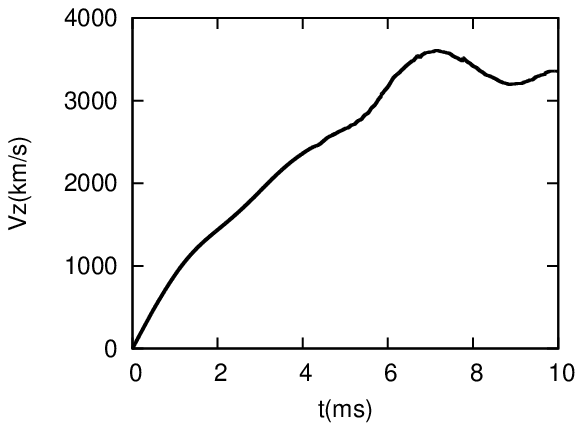}
\plotone{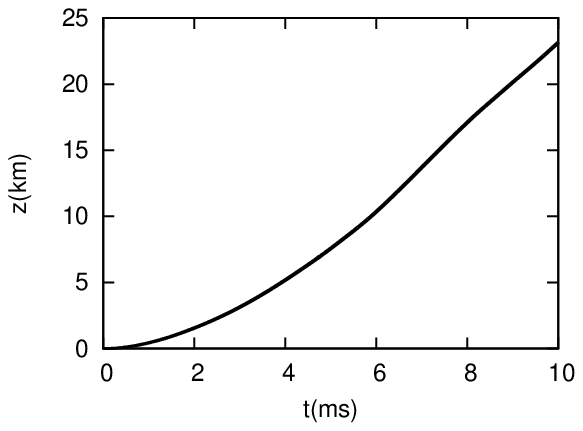}
\plotone{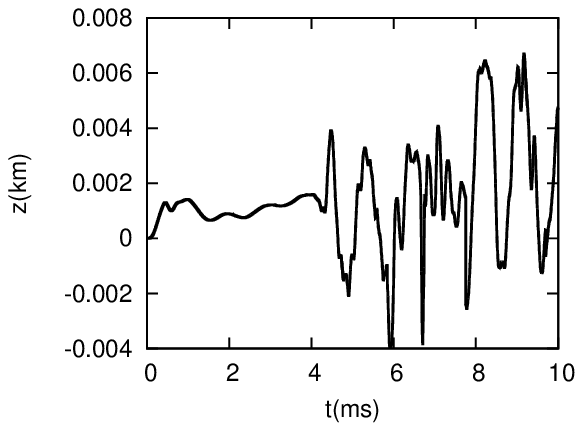}
\caption{Same as Figure~\ref{fig:Oscivelo} but for the runaway of PNS.
\label{fig:Kickvelo}} 
\end{figure}

In realistic simulations the shock wave expands asymmetrically after successful shock revival and expels the envelope anisotropically. The PNS then undergoes a recoil \citep{1994A&A...290..496J,2012MNRAS.423.1805N}. In this section, we mimic such runaways of PNS very crudely. The initial configurations of matter and neutrinos are the same as those used in the previous section. In this case, however, we do not add velocity perturbations. Instead, we continuously add by hand
the external acceleration of $10^{11} {\rm cm}/{\rm s^2}$ in the positive z-direction within the region of $r < 30$km on the moving grid. The simulation was carried out for $t=10$ms.

Figure~\ref{fig:Kickvelo} is the counter part of Figure~\ref{fig:Oscivelo} for the present case. As expected, the PNS moves continuously in the positive z-direction. It is seen in the top panel that the PNS velocity reaches $\sim 3000$km/s at the end of the simulation, which is much larger than the realistic kick velocity of a few hundred km/s. In spite of this rather extreme runaway of PNS, the moving-mesh tracks it very well, as shown in the lower panel of the figure. In fact, the distance between the mass center of PNS and the origin of the moving-mesh remains less than $10^{-2}$km, which is close enough to avoid numerical problems. Incidentally, oscillations that are evident after $t \sim 4$ms in the lower panel are ascribed to the deformations of PNS.

\begin{figure*}
\vspace{15mm}
\epsscale{1.0}
\plottwo{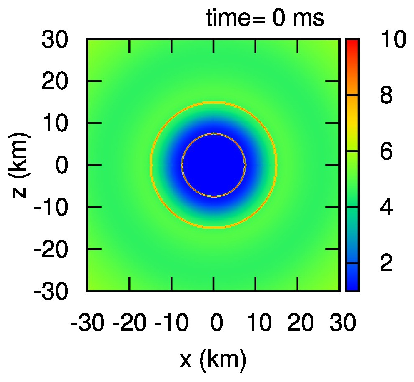}{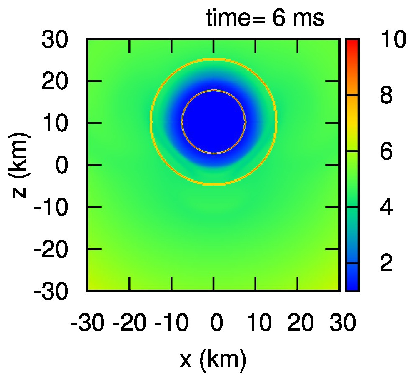}
\plottwo{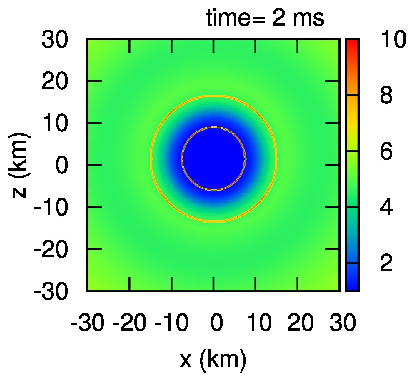}{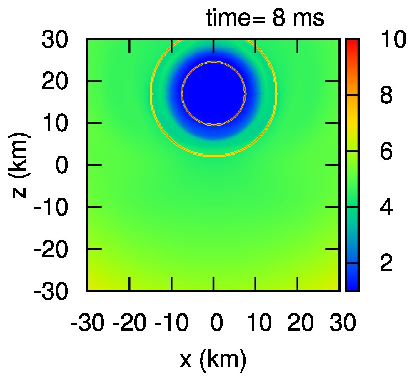}
\plottwo{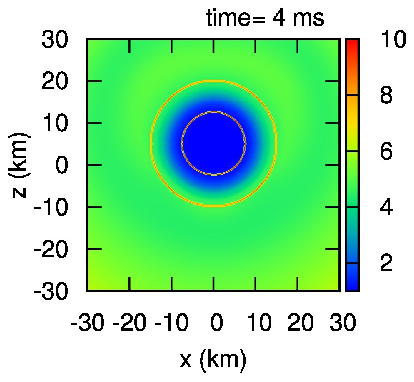}{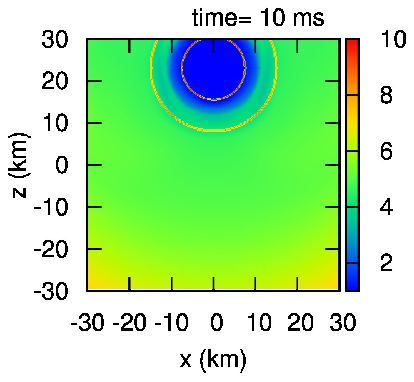}
\caption{A series of entropy contours on the fixed coordinates for runaway of PNS. The bluish regions of low entropies correspond to the un-shocked core of PNS. The two red circles are concentric to the origin of the moving-mesh.
\label{fig:entrocontKick}} 
\end{figure*}

\begin{figure*}
\vspace{15mm}
\epsscale{1.0}
\plottwo{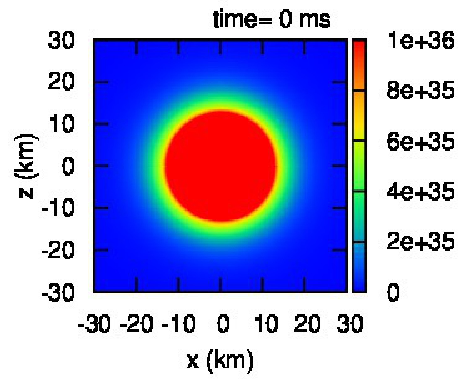}{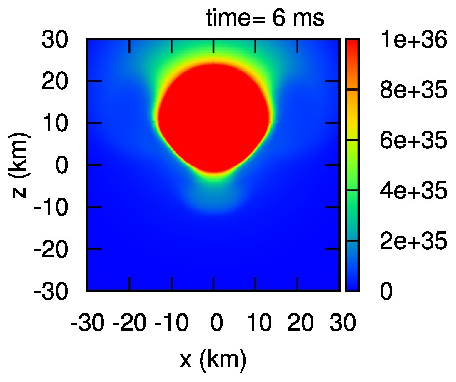}
\plottwo{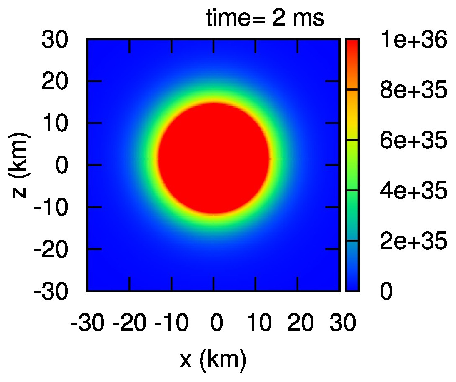}{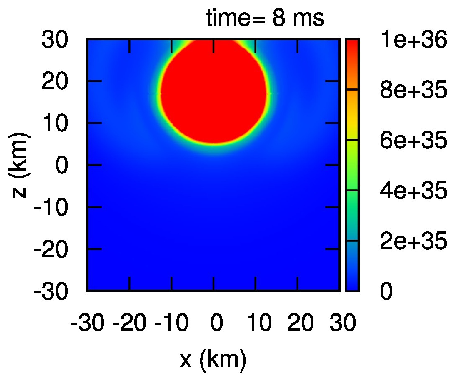}
\plottwo{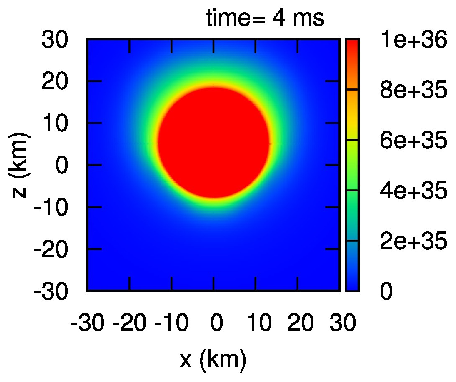}{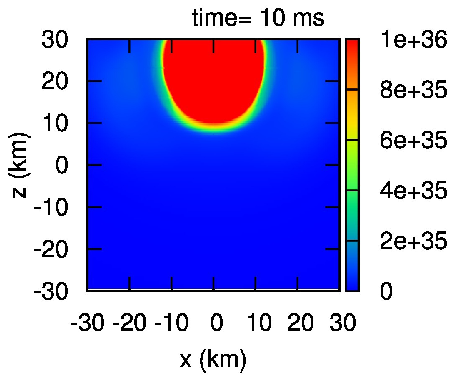}
\caption{Same as Figure~\ref{fig:entrocontKick} but for the number density of electron-type neutrinos.
\label{fig:NuenumcontKick}} 
\end{figure*}

A series of snapshots of the entropy distribution in the meridian plane are shown in Figure~\ref{fig:entrocontKick}. The low entropy region with deep blue colors in this figure corresponds to the central un-shocked part of PNS. It is apparent that the PNS moves upward with time. Note that this figure is drawn on the fixed coordinates, which coincide with the moving coordinates initially. In each panel of the same figure, we put two concentric circles, which represent the moving-mesh. It is confirmed again that they trace the PNS closely. In Figure~\ref{fig:NuenumcontKick}, on the other hand, the (number) density contour is drawn in color for electron-type neutrinos. One can see that the neutrino density is slightly non-spherical owing to the deformation of PNS. It is more important, however, that the neutrinos are comoving with PNS. This is of course a consequence of the neutrino trapping, which occurs in the optically thick region. In the Boltzmann-Hydro simulation, however, it is highly non-trivial and is in fact ensured by the combination of the following two conditions: (1) neutrinos are isotropically distributed in the fluid-rest frame; (2) the neutrino distribution in the O-frame is related accurately with that in the fluid-rest frame by a Lorentz transformation. The result we have just presented is yet another demonstration that our code is working properly.

Last but not least, we mention the conservation of linear momentum in our code. It is a well known fact that it is difficult in general for hydrodynamics codes like ours that adopt curvilinear coordinates to enforce the conservation of linear momentum. It is evident indeed in Eqs~(\ref{eq:Mon3pra1}) or (\ref{eq:Wh}) that these equations can not be written in the conserved form even in the absence of gravity and neutrino interactions. Not to mention, the gravity term written not in the conservation form also attributes to the violation of momentum (and energy) conservation (see also \citet{2010ApJS..189..104M}). Note, however, that as formulated in Section~\ref{subsec:shiftvector}, we do not use the conservation law to evaluate the PNS velocity and its acceleration. Regardless, we checked quantitatively the violation of linear momentum in our code by conducting another 2D purely hydrodynamical simulation (same initial conditions as previous tests but adding 1$\%$ random density perturbation) for 100ms. We found that the numerical error is equivalent to $\sim 10$km/s of the PNS's kick velocity, which is not negligibly small but still much smaller than the typical velocity of $\sim 100$km/s. Considering the rather coarse grid employed in this paper and the purpose of this paper, we may conclude that our code performed well.

\section{Summary} \label{sec:summary}

In this paper, we have presented a novel method to deal with motions of PNS on spherical polar coordinates. It is based on a moving-mesh technique, as far as the neutrino transport part is concerined, it is essentially equivalent to the general relativistic extension of the special relativistic Boltzmann solver we developed earlier. In fact, the Boltzmann equation is reformulated in the 3+1 formalism of GR although the GR code thus obtained is applied only to the flat spacetime coupled with the Newtonian hydrodynamics code and self-gravity module in this paper.
 The shift vector is utilized to specify the movement of spatial coordinates so that they could track the PNS motion approximately. As a matter of fact, without such a technique we encountered a numerical crash with unphysical features emerging at the coordinate origin, as the PNS is dislocated from the original position. Since the coordinate origin stays very close to the mass center of PNS with the moving-mesh technique, we expect that in more realistic simulations the extended Boltzmann-Hydro code will be able to treat the violent oscillations and ultimate runaway of PNS in the post-bounce phases of CCSNe on spherical polar coordinates.

 We have also described in detail the numerical implementations of the GR extensions to our SR Boltzmann code, which was constructed on two energy grids so that it could deal with both the advection and collision terms correctly. It turns out that the extension is rather straightforward thanks to the use of appropriate tetrads. The two energy grids and the transformations between them, which are employed in the SR Boltzmann code, are nicely identified with these tetrads and their Lorentz transformations, respectively.

 In Section~\ref{sec:twodsim}, we have validated our method by applying it to two test problems: toy models of PNS oscillations and a runaway, which mimic very crudely more realistic post-bounce simulations. We have demonstrated that the code has nice tracking capabilities and can follow the evolutions without any problems such as those we encountered when the fixed spatial grid was deployed. Incidentally, the extended code is currently being applied to realistic 2D simulations of CCSNe and the results and a thorough code validation will be reported elsewhere soon. We are also planning to conduct truly GR Boltzmann simulations of neutrino transport in a black hole spacetime. It should be apparent that any (local) gauge conditions can be imposed in addition to the uniform shift we employed in this paper.

The next step is to couple the GR neutrino transport code with a solver of the Einstein equations. Note that the hydrodynamics code is already GR (see \citet{2008ApJ...689..391N,2009ApJ...696.2026N}) although the Newtonian version was used in this paper. Such an integrated code, once completed, will certainly broaden the scope of application much beyond CCSNe.


\acknowledgments 
We are grateful to A. Juodagalvis for providing the data of electron capture rates on heavy nuclei. H.N. acknowledges to M. Shibata, Y. Sekiguchi and H. Okawa for valuable comments and discussions. H.N. also thanks Werner Marcus for proofreadings. The numerical computations were performed on the supercomputers at K, at AICS, FX10 at Information Technology Center of Tokyo University, SR16000 at YITP of Kyoto University, and SR16000 at KEK under the support of its Large Scale Simulation Program (14/15-17, 15/16-08), Research Center for Nuclear Physics (RCNP) at Osaka University. Large-scale storage of numerical data is supported by JLDG constructed over SINET4 of NII. H.N. was supported in part by JSPS Postdoctoral Fellowships for Research Abroad No. 27-348. This work was also supported by Grant-in-Aid for the Scientific Research from the Ministry of Education, Culture, Sports, Science and Technology (MEXT), Japan (15K05093, 24103006, 24740165, 24244036, 25870099) and HPCI Strategic Program of Japanese MEXT and K computer at the RIKEN (Project ID: hpci 130025, 140211, and 150225).

\end{document}